\documentclass[aps,prl,english,amsmath,floatfix,amssymb,superscriptaddress,tightenlines,twocolumn,nofootinbib]{revtex4-2}
\usepackage{hyperref}
\usepackage{mdframed}
\usepackage{graphicx}
\usepackage{amsmath}
\usepackage{tikz}
\usepackage{soul,xcolor}
\usepackage{amssymb}
\usepackage{lipsum, babel}
\usepackage{amsthm}
\usepackage[shortlabels]{enumitem}
\usepackage{tikz-cd}
\usepackage{breakurl}

\setstcolor{red}
\usetikzlibrary{positioning}
\usetikzlibrary{patterns}
\usetikzlibrary{arrows.meta}
\usetikzlibrary{spy}

\makeatother

\theoremstyle{definition}

\theoremstyle{theorem}

\theoremstyle{corollary}

\theoremstyle{lemma} 

\theoremstyle{Proposition}

\definecolor{IKblue}{RGB}{25,25,125}
\definecolor{BSorange}{RGB}{140,50,0}

\newcommand{\axiomone}[2] 
{
	\begin{scope}[xshift=#1cm, yshift=#2cm]
		\draw[line width=0.6pt, fill=red!10!white, opacity=0.5] (0,0) circle (0.3cm);
		\draw[line width=0.6pt] (0,0) circle (0.3cm);
		\draw[line width=0.6pt] (0,0) circle (0.15cm);
	\end{scope}
}

\newcommand{\Rom}[1]{\uppercase\expandafter{\romannumeral #1\relax}}

\newcommand{\calC}{{\mathcal C}}
\newcommand{\calD}{{\mathcal D}}

\newcommand{\Tr}{{\rm Tr}}

\begin{document}

\title{Domain wall topological entanglement entropy} 	
\author{Bowen Shi}
\affiliation{Department of Physics, University of California at San Diego, La Jolla, CA 92093, USA}
\affiliation{Department of Physics, The Ohio State University, Columbus, OH 43210, USA}

\author{Isaac H. Kim}
\affiliation{3 Centre for Engineered Quantum Systems, School of Physics, University of Sydney, Sydney, NSW 2006, Australia}

\date{\today}
		
\begin{abstract}
We study the ground-state entanglement of gapped domain walls between topologically ordered systems in two spatial dimensions. We derive a universal correction to the ground-state entanglement entropy, which is equal to the logarithm of the total quantum dimension of a set of superselection sectors localized on the domain wall. This expression is derived from the recently proposed entanglement bootstrap method.
\end{abstract}
\maketitle

Topological order is a new kind of order that lies outside of Landau's symmetry breaking paradigm~\cite{Wen2004}. This refers to a phase of matter that exhibits exotic phenomena such as topology-dependent ground-state degeneracy~\cite{Wen1990} and fractional statistics~\cite{Leinaas1977,PhysRevLett.49.957}. A systematic understanding of these phenomena is one of the fundamental goals in physics. More practically, such systems may pave ways to build a fault-tolerant quantum computer~\cite{Kitaev2003}.

When two topologically ordered systems are joined together along their boundaries, one may obtain a \emph{gapped domain wall} between the two~\cite{Bravyi1998,Beigi2010,KitaevKong2012}. Gapped domain walls can lead to novel phenomena, such as the change in the ground-state degeneracy~\cite{Bravyi1998,Hung2015,Lan2015} and the emergence of superselection sectors associated with point-like excitations that are not uniquely determined by the bulk data~\cite{KitaevKong2012,Kong2014}. 

While there has been a flurry of recent work dedicated to gapped domain walls, one aspect of it remains unknown. Can we detect gapped domain walls from ground-state entanglement? In the bulk of topologically ordered systems, there is a universal correction to the entanglement entropy over a disk-like region that reveals nontrivial information about the underlying topological phase~\cite{Kitaev2006,Levin2006}, given by the total quantum dimension of the anyonic excitations. Moreover, codimension-$2$ defects give rise to an extra universal correction~\cite{Brown2013}. However, whether gapped domain walls give rise to such universal contribution has remained open. If such a contribution exists, what would be its physical meaning?

In this letter, we provide a definitive answer to these questions. Specifically, we study the entanglement entropy -- defined as $S_A = -\Tr(\sigma_A \ln \sigma_A)$ with respect to a global state $\sigma$ -- over a subsystem $A$ that the domain wall passes through. Our main finding is that there is a universal contribution to the entanglement entropy that reveals nontrivial information about the property of the domain wall.
 
\begin{figure}[h]
    \centering
    \includegraphics[width=0.95\columnwidth]{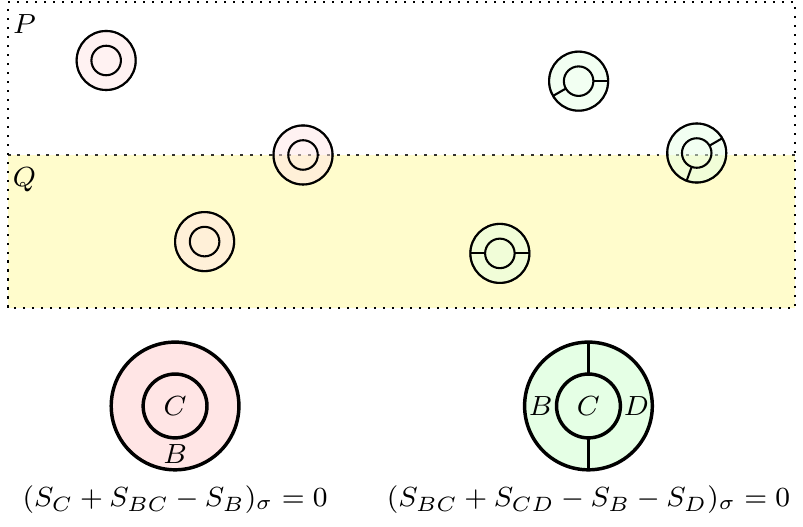}
    \caption{Summary of our assumptions. We assume that $(S_C + S_{BC} - S_B)_{\sigma}=0$ (red) and $(S_{BC} + S_{CD} - S_B - S_D)_{\sigma}=0$ (green), both in the bulk and on the domain wall, over arbitrarily large regions that can be smoothly deformed from the shown configurations.  Here, $(\ldots)_{\sigma}$ means that the entanglement entropies appearing in the parentheses is computed with respect to the reference state $\sigma$.  The subsystems are allowed to be deformed as long as the boundaries between $B$ and $D$ do not cross the domain wall.}
    \label{fig:assumptions}
\end{figure}

To be concrete, consider two topologically ordered mediums in two spatial dimensions, denoted as $P$ and $Q$, that are separated by a domain wall; see Fig.~\ref{fig:assumptions}. We will assume that the ground state of this system is well approximated by a quantum state $\sigma$ that obeys the assumptions we soon describe below. While we expect these assumptions to hold in ground states of gapped systems, at least approximately, our analysis depends solely on the assumptions imposed on $\sigma$. Therefore, instead of referring to $\sigma$ as the ground state, we will refer to it as the \emph{reference state}.

The reference state is assumed to obey the assumptions summarized in Fig.~\ref{fig:assumptions}.\footnote{While the assumptions in Fig.~\ref{fig:assumptions} concern arbitrarily large regions, one can verify these assumptions locally. Specifically, if these assumptions hold on every ball of bounded radius, they continue to do so on arbitrarily large scales \cite{EntanglementBootstrap_long}. Therefore, in principle, given sufficiently many copies of the quantum state $\sigma$, one can verify this condition directly, in a time that scales merely linearly with the system size.
} While these assumptions will generally hold only approximately, in particular, in models that are away from a fixed point of a renormalization group flow, we believe our conclusion is applicable to those models as well. The main reason is that the key technical statements we use rest on a general fact about tripartite quantum states: that if a tripartite state $\rho_{ABC}$ satisfies an entropy identity $(S_{AB} + S_{BC} - S_B - S_{ABC})_{\rho}=0$, the state has the following \emph{Markov chain structure}:
\begin{equation}
    \rho_{ABC} =  \Phi_{B\to AB} [\rho_{BC}] \label{eq:local_recovery}
\end{equation}
for some quantum channel $\Phi_{B\to AB}$ acting on $B$, which sends density matrices in $B$ to density matrices in $AB$. It is well known that if the entropy identity holds \emph{approximately}, \emph{i.e.,} $(S_{AB} + S_{BC} - S_B - S_{ABC})_{\rho}\approx 0$, then so does Eq.~\eqref{eq:local_recovery}~\cite{2015CMaPh.340..575F}. 



\begin{figure}[h]
	\centering
	\includegraphics[width=0.9\columnwidth]{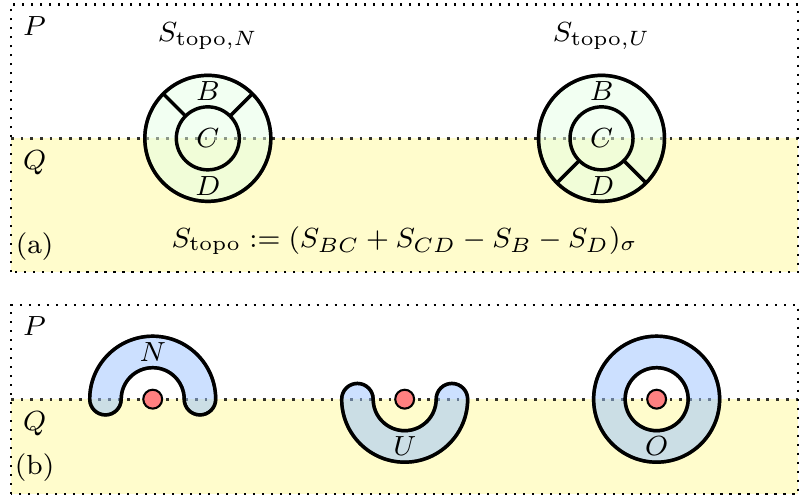}
	\caption{(a) Subsystems involved in the definition of the domain wall topological entanglement entropy. (b) Three ways to detect a point-like excitation (red dot) on the domain wall. Measurement processes are considered in the blue regions. The parton sectors can be detected from a measurement process strictly localized on $N$-shaped (left) and $U$-shaped (middle) subsystems. In comparison, the set of superselection sectors of point-like excitations in the vicinity of the gapped domain wall can be detected from a measurement on $O$-shaped (right) subsystems.}
	\label{fig:tee_configs}
\end{figure}

Surprisingly, from these seemingly minimalistic assumptions and observations, we can deduce the universal properties of the underlying quantum phase. Of particular importance to us is the existence of superselection sectors called \emph{parton  sectors}~\cite{EntanglementBootstrap_long}. These sectors subdivide the known superselection sectors of the point-like excitations on the domain wall~\cite{KitaevKong2012,Kong2014}. Furthermore, they represent the complete set of topological charges that can be measured from the $N$- and  $U$-shaped regions in Fig.~\ref{fig:tee_configs}. Intuitively, when the red dot in Fig.~\ref{fig:tee_configs}(b) is obtained by fusing an anyon from the $P$ ($Q$) phase to the wall, it may result in a nontrivial $N$-type ($U$-type) parton sector. 

Furthermore, our assumptions allow us to define and derive an exact expression for the domain wall analog of the topological entanglement entropy~\cite{Kitaev2006,Levin2006}. 
The underlying method, which we refer to as \emph{entanglement bootstrap}, has recently been initiated and is under development in various contexts~\cite{Kim2015sydney,SKK2019,2020PhRvR...2b3132S,EntanglementBootstrap_long}.

In this letter, we will focus on the derivation of the  \emph{domain wall topological entanglement entropies}, introduced below. We define two quantities,  $S_{\text{topo}, N}$ and $S_{\text{topo}, U}$, which are linear combinations of entanglement entropies of $\sigma$ over the subsystems described in Fig.~\ref{fig:tee_configs}(a). We derive the following expressions:
\begin{equation}
\begin{aligned}
S_{\text{topo}, N} &= 2\ln \mathcal{D}_N, \\
S_{\text{topo}, U} &= 2\ln \mathcal{D}_U,
\end{aligned}
    \label{eq:main_result}
\end{equation}
where $\mathcal{D}_N$ and $\mathcal{D}_U$ are the \emph{total quantum dimensions} of the $N$- and $U$-type parton sectors:
\begin{equation}
    \begin{aligned}
    \mathcal{D}_N=\sqrt{\sum_{n\in \calC_N} d_n^2} \quad \text{ and }  \quad
    \mathcal{D}_U=\sqrt{\sum_{u\in \calC_U} d_u^2}.
    \end{aligned}\label{eq:total_qd}
\end{equation}
Here we have denoted the set of $N$-type ($U$-type) parton sectors as $\calC_N=\{ 1, n, \cdots\}$ ($\calC_U=\{1, u,\cdots \}$); $d_n$ and $d_u$ are the \emph{quantum dimensions} of  the parton sectors, which shall be defined later in Eq.~\eqref{eq:quantum_dimension_parton}.

One can gain intuition about the parton sectors from some examples. For instance, consider a gapped domain wall that separates the toric code~\cite{Kitaev2003} from a product state. If we impose the electric boundary condition~\cite{Bravyi1998}, the set $\calC_N$ contains two Abelian sectors and $\calC_U$ contains a unique (vacuum) sector. 
However, there can be parton sectors with quantum dimensions strictly larger than $1$. For example, one of the gapped domain wall types~\cite{Beigi2010} between the non-Abelian $S_3$ quantum double (on the $P$ side) and the toric code (on the $Q$ side) has $\calC_N=\{ 1, n\}$ and $\calC_U=\{ 1\}$, where $d_n=\sqrt{2}$. 

The parton sectors can be detected from far away, by presumably performing an Aharonov-Bohm type interference experiment, but by merely closing a ``half-loop''; see Fig.~\ref{fig:tee_configs}(b). That the different sectors can be discerned this way suggests that there is a profound form of ground-state entanglement in the vicinity of the domain wall that cannot be probed locally. Domain wall topological entanglement entropy precisely captures this nonlocal piece of information. In the remainder of this letter, we shall justify this physical interpretation. 


\emph{Information convex sets.---}
Let us begin with the essential concept called information convex set~\cite{SKK2019}; see also~\cite{Kim2015sydney,Shi2018,2018arXiv180101519S}.
For the readers' convenience, we will use a definition of information convex set, which is different from its original form~\cite{SKK2019}. Because these two definitions are equivalent under the assumptions in Fig.~\ref{fig:assumptions}, we do not lose any generality in our argument. The advantage of the new definition is that it simplifies many of our analyses, facilitating us in focusing on the essential ideas.

Consider a set of microscopic degrees freedom arranged on a plane, denoted as $\Lambda$. Let  $\Omega\subset \Lambda$ be a smooth subsystem, \emph{e.g.,} a disk or an annulus. The information convex set $\Sigma(\Omega)$, for a given reference state $\sigma$, is the set of density matrices on $\Omega$ that satisfies the following property. For any $\rho_{\Omega}\in \Sigma(\Omega)$ and any $\Omega' \supset \Omega$ obtained by expanding $\Omega$ along its boundary while retaining its topology, there is a density matrix $\rho'_{\Omega'}$ such that (1) $\rho'_{\Omega'}$  is indistinguishable from the reference state $\sigma$ over every disk-like region contained in $\Omega'$ and (2) $\Tr_{\Omega'\setminus \Omega} \, \rho'_{\Omega'} = \rho_{\Omega}$. 

The information convex set $\Sigma(N)$ has an intimate connection with $S_{\text{topo}, N}$, where $N$ is the $N$-shaped region depicted in Fig.~\ref{fig:tee_configs}(b). 
Specifically, suppose $\Sigma(N)$ contains more than one element. Then we must conclude that $S_{\text{topo}, N}$ is nonzero. 

\begin{figure}[h]
	\centering
	\includegraphics[width=0.9\columnwidth]{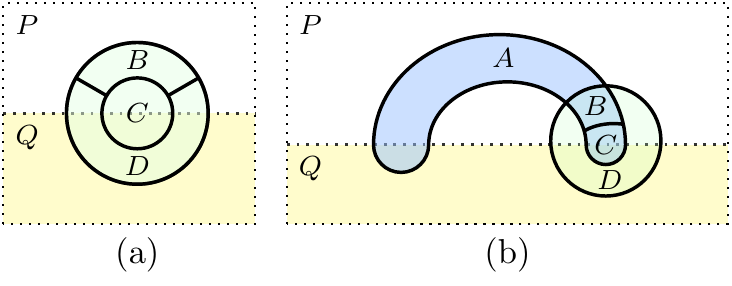}
	\caption{(a) Subsystems relevant to the domain wall topological entanglement entropy $S_{\text{topo}, N}$. (b) A partition of $N$ into $ABC$. $BCD$ is a disk.}
	\label{fig:domain_wall_axiom}
\end{figure}

To understand why, it is helpful to assume that $S_{\text{topo}, N}$ vanishes and study the consequence of this assumption. Consider a partition of $N$ into $ABC$ in Fig.~\ref{fig:domain_wall_axiom}(b). By using the property of the information convex set, we can extend the state $\rho_{N}$ to some state $\rho'_{ND}$, where $BCD$ is a partition of a disk topologically equivalent to that in Fig.~\ref{fig:domain_wall_axiom}(a). With these subsystems, we get
\begin{equation}
\begin{aligned}
    I(A:C\vert B)_{\rho} &\leq (S_{BC} + S_{CD} - S_B - S_D)_{\rho'} \\
    &= (S_{BC} + S_{CD} - S_B - S_D)_{\sigma}\\
    &=0
\end{aligned}
    \label{eq:markov_N}
\end{equation}
for any $\rho_N \in \Sigma(N)$, where $I(A:C|B)_{\rho} := (S_{AB} + S_{BC} - S_B - S_{ABC})_{\rho}$ is the \emph{conditional mutual information}. The first line follows from the strong subadditivity of entropy (SSA)~\cite{Lieb1973}. In the second line, we used the fact that $\rho'$ is indistinguishable from $\sigma$ on the disk $BCD$. Equation~\eqref{eq:markov_N} implies that $\rho_N$ is uniquely determined by $\rho_{AB}$ and $\rho_{BC}$, which are equal to $\sigma_{AB}$ and $\sigma_{BC}$ respectively~\cite{Kim2014a}. Therefore, if $S_{\text{topo}, N}=0$, $\Sigma(N)$ must contain a unique element, proving our claim.

\emph{Structure of $\Sigma(N)$.---}
While information convex sets generally do not have a particularly noteworthy structure, they become highly constrained if the reference state obeys the assumptions in Fig.~\ref{fig:assumptions}. Of particular importance to us is the subsystem $N$ in Fig.~\ref{fig:tee_configs}. The information convex set of $N$ forms a simplex:
\begin{equation}
    \Sigma(N) = \left\{  \bigoplus_n p_n \rho^n_N  :  \sum_n p_n=1, p_n \ge 0 \right \},\label{eq:simplex_parton_n}
\end{equation}
where the extreme points $\rho_N^n$ are mutually orthogonal to each other~\cite{EntanglementBootstrap_long}. The set of labels can be identified with the set of $N$-type parton sectors $\mathcal{C}_N =\{1,n, \dots \}$, where $1$ denotes the vacuum sector associated with the extreme point $\rho_N^1=\sigma_N$. A similar conclusion holds for the subsystem $U$ in Fig.~\ref{fig:tee_configs}. 

From the argument above, we see that the existence of a nontrivial parton sector implies a nonzero value of domain wall topological entanglement entropy. A sharper constraint on the topological entanglement entropy can be obtained in terms of the \emph{quantum dimensions} of the parton sectors. 
In our theory~\cite{EntanglementBootstrap_long}, the quantum dimension for a parton sector $n\in \calC_N$ is defined as
\begin{equation}
    d_n := \exp \left( \frac{S(\rho_N^n) - S(\rho_N^1)}{2}\right). \label{eq:quantum_dimension_parton}
\end{equation}
While the quantum dimension may appear to depend on the choice of the underlying subsystem $N$, it does not; the right-hand side of Eq.~\eqref{eq:quantum_dimension_parton} is invariant under smooth deformations of $N$, justifying our notation~\cite{EntanglementBootstrap_long}. See the Appendix for a review of the identities that the quantum dimensions must satisfy, thus justifying our definition.


Using Eq.~\eqref{eq:quantum_dimension_parton} and the orthogonality of the extreme points, we can quantify the ``maximal ignorance" about the parton sector. Let ${\tau}_N$ be the maximum-entropy element of $\Sigma(N)$, then
\begin{equation} \label{eq:ent_diff}
\begin{aligned}
S({\tau}_N) - S(\sigma_N) &= \max_{\{p_n\}} S\left(\sum_n p_n \rho^n_N \right) - S(\sigma_N)\\
 &=\max_{\{p_n\}} \left(H\left(\{p_n \}\right) + \sum_{n\in \mathcal{C}_N} p_n \ln d_n^2 \right)\\
 &= 2\ln \mathcal{D}_N,
\end{aligned} 
\end{equation}
where $\{p_n\}$ is a probability distribution and $H(\{ p_n\}) :=  - \sum_n p_n \ln p_n $. To obtain the third line, we have plugged in the optimal choice of $p_n = d_n^2/\calD_N^2 $. We may interpret the result in Eq.~(\ref{eq:ent_diff}) as the maximal ignorance of the parton sector; this quantifies the amount of information that remain unspecified if we only have access to subsystems $AB$ and $BC$.

The conditional mutual information is lower bounded by this maximal ignorance:
\begin{equation}
\begin{aligned}
I(A:C|B)_{\sigma} & = S({\tau}_N) - S(\sigma_N) +I(A:C|B)_{{\tau}}\\
& \ge 2\ln \mathcal{D}_N. \label{eq:cmi_lower_bound}
\end{aligned}
\end{equation}

\begin{figure}[h]
	\centering
	\includegraphics[width=0.9\columnwidth]{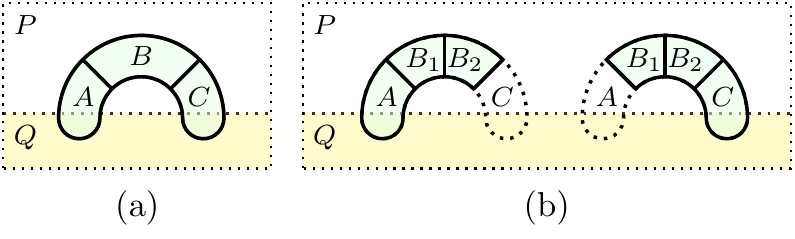}
	\caption{(a) Partition of $N=ABC$. (b) Further partition $B=B_1 B_2$ and subsystems relevant to merging.}
	\label{fig:ABC}
\end{figure}

\emph{Merging density matrices.---}
Remarkably, Eq.~\eqref{eq:cmi_lower_bound} actually holds with an \emph{equality}. This identity follows from the following important property of the maximum-entropy element ($\tau_N=\sum_n \frac{d_n^2}{\mathcal{D}_N^2} \rho_{N}^n$) of $\Sigma(N)$:
\begin{equation}
    S(\tau_{N}) = S(\sigma_{AB}) + S(\sigma_{BC}) - S(\sigma_B).
\end{equation}

We can establish this fact using the \emph{merging technique}~\cite{Kato2016,SKK2019}. Specifically, consider density matrices $\rho_{AB_1B_2}$ and $\lambda_{B_1B_2C}$ such that $I(A:B_2|B_1)_{\rho}=I(B_1:C|B_2)_{\lambda}=0$ and $\rho_{B} = \lambda_{B}$, where $B=B_1B_2$. Then there exists a density matrix $\widetilde{\tau}_{ABC}$ such that $\widetilde{\tau}_{AB} = \rho_{AB}$ and $\widetilde{\tau}_{BC}=\lambda_{BC}$. Moreover, $\widetilde{\tau}_{ABC}$ obeys
\begin{equation}
I(A:C|B)_{\widetilde{\tau}}=0,\label{eq:tau_cmi}
\end{equation}
which means that $\widetilde{\tau}_{ABC}$ is the {unique} maximum-entropy state consistent with $\rho_{AB}$ and $\lambda_{BC}$~\cite{Kato2016}. Here we say two states are consistent with each other if their reduced density matrices on their overlapping support are identical.

We can use this result in the following way. Partition $N$ into $A,$ $B=B_1B_2$, and $C$, as is shown in Fig.~\ref{fig:ABC}(b). It follows from our assumption that
\begin{equation}
    I(A:B_2|B_1)_{\sigma} = I(B_1:C|B_2)_{\sigma}=0,
\end{equation}
using the same logic that led to Eq.~\eqref{eq:markov_N}. Moreover, $\sigma_{AB}$ and $\sigma_{BC}$ are identical on $B$ because they are obtained from the same reference state. Therefore, there exists a unique state $\widetilde{\tau}_{ABC}$ consistent with both $\sigma_{AB}$ and $\sigma_{BC}$ that satisfies Eq.~\eqref{eq:tau_cmi}.

The fact that $\widetilde{\tau}_{ABC}$ is consistent with $\sigma_{AB}$ and $\sigma_{BC}$ suggests that $\widetilde{\tau}_{ABC}$ may belong to $\Sigma(N)$. This turns out to be correct, provided that the involved subsystems are sufficiently large.\footnote{Specifically, $A$ and $C$ must be separated by a distance large compared to the radius of the minimal disk that obeys the conditions in Fig.~\ref{fig:assumptions}. This fact was rigorously established in Ref.~\cite{SKK2019}; see Section II of Ref.~\cite{EntanglementBootstrap_long} for a review.} 
Thus, the maximum-entropy state of $\Sigma(N)$ is identical with the merged state, $\tau_{ABC}=\widetilde{\tau}_{ABC}$, and therefore $I(A:C\vert B)_{\tau}=0$. 
This implies that the lower bound in Eq.~\eqref{eq:cmi_lower_bound} saturates, leading to the following expression:
\begin{equation}
    I(A:C|B)_{\sigma} = 2\ln \calD_N. \label{eq:tee_expression_N}
\end{equation}
By Eq.~\eqref{eq:ent_diff}, Eq.~\eqref{eq:tee_expression_N} quantifies the maximal ignorance about the parton sector, given access to subsystems $AB$ and $BC$ in Fig.~\ref{fig:ABC}(a).



\begin{figure}[h]
	\centering
	\includegraphics[width=0.65\columnwidth]{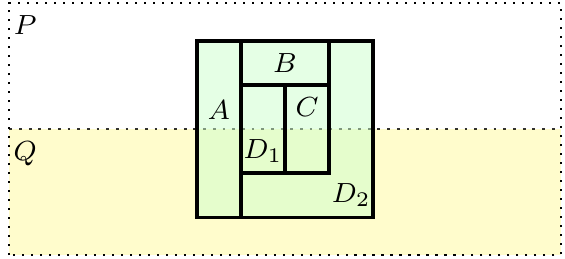}
	\caption{The partition used in the proof of Eq.~\eqref{eq:main_result}. $D_1D_2=D$.}
	\label{fig_proof_complicated}
\end{figure}

\emph{Domain wall topological entanglement entropy.---}
Next, let us establish the equivalence of $I(A:C|B)_{\sigma}$ to the domain wall topological entanglement entropy $S_{\text{topo}, N}$. 
Consider the partition in Fig.~\ref{fig_proof_complicated}, which contains both the region $ABC$ used in Eq.~\eqref{eq:tee_expression_N} and the $BCD$ for the definition of $S_{\text{topo}, N}$. The following is an important identity:
\begin{equation}
\begin{aligned}
&S_{\text{topo}, N} - I(A:C|B)_{\sigma}\\
=& (S_{ABC} + S_{CD} - S_D - S_{AB})_{\sigma}.
\end{aligned}
\label{eq:1035}
\end{equation}
We show that the bottom line of Eq.~\eqref{eq:1035} vanishes, thus establishing $S_{\text{topo}, N} = I(A:C|B)_{\sigma}$. This can be shown by lower and upper bounding $(S_{ABC} + S_{CD} - S_D - S_{AB})_{\sigma}$ by $0$. Note that $(S_{ABC} + S_{CD} - S_D - S_{AB})_{\sigma}\ge 0$ follows straightforwardly from SSA. Moreover,
\begin{equation}
\begin{aligned}
&(S_{ABC} + S_{CD} - S_D - S_{AB})_{\sigma}\\
\le & (S_{ABCD_1} + S_{CD} - S_{D_2} - S_{AB})_{\sigma}\\
=&0.
\end{aligned}
\end{equation} 
The second line follows from SSA. The third line is obtained by applying our assumptions in Fig.~\ref{fig:assumptions}.

Therefore, $I(A:C|B)_{\sigma} = S_{\text{topo}, N}$. Of course, the same analysis applies to $S_{\text{topo}, U}$. This leads to our main conclusion:
\begin{equation}
    S_{\text{topo}, N} = 2 \ln \calD_N \quad \textrm{and} \quad  S_{\text{topo}, U} = 2 \ln \calD_U.
\end{equation}

\emph{Summary.---} We have proposed a domain wall analog of topological entanglement entropy and derived its exact expression. We envision this to be a valuable tool to detect the presence of nontrivial gapped domain walls from ground-state entanglement.

\emph{Acknowledgments.--- } The work of I. K. was supported by the Simons Foundation It from Qubit Collaboration and by the Australian Research Council via the Centre of Excellence in Engineered Quantum Systems (EQUS) Project No. CE170100009. B. S. is supported by the National Science Foundation under Grant No. NSF DMR-1653769, University of California Laboratory Fees Research Program, grant LFR-20-653926, as well as the Simons Collaboration on Ultra-Quantum Matter, Grant No. 651440 from the Simons Foundation.

\appendix
\section{Appendix: quantum dimension}
We provide a brief overview of the parton sectors, with the primary emphasis on justifying our definition of \emph{quantum dimension} Eq.~\eqref{eq:quantum_dimension_parton}.
 While the following discussion already appeared in Ref.~\cite{EntanglementBootstrap_long}, we provide a more succinct explanation behind our choice, drawing analogy with the well-known theory of anyon. 

An anyon theory is equipped with a set of superselection sectors $\mathcal{C}= \{1, a, b, c, \ldots \}$, where $1$ is the vacuum sector. The set of quantum dimensions of these sectors, denoted as $\{d_a: a\in \mathcal{C} \}$ are positive numbers uniquely determined by the following relation: 
\begin{equation}
	d_a d_b = \sum_c N_{ab}^c d_c,\label{eq:fusion_anyon}
\end{equation}
where $N_{ab}^c$ is the dimension of the fusion space in which the anyon $a$ and $b$ fuse into $c$.
The key point is that the quantum dimensions can be uniquely determined given a set of (integer-valued) fusion space dimensions; see Appendix E of Ref.~\cite{Kitaev2006solo}.

The quantum dimensions of the parton sectors can be uniquely determined by fusion space dimensions as well. When the anyon content of one side of the domain wall is trivial, the quantum dimensions of parton sectors obey the following analogous equation~\cite{EntanglementBootstrap_long}:
\begin{equation}
	d_n d_m = \sum_{p} N_{nm}^p d_p, \label{eq:fusion_parton}
\end{equation}
where $N_{nm}^p$ is the dimension of the fusion space in which the parton $n$ and $m$ fuse into a parton $p$, where $n,m,p \in \calC_N$. This fact suggests that our definition of quantum dimension is a sensible one.

However, more generally, when both sides of the domain wall have nontrivial anyon contents, the condition that defines the quantum dimensions will be different from Eq.~(\ref{eq:fusion_parton}).\footnote{This is because, in general, there is no well-defined fusion space involving three partons, and the integers $N_{nm}^p$ are no longer defined.}
In this case, the quantum dimensions of parton sectors are determined by \emph{two} sets of fusion space dimensions $\{N_{\mu(n)}^{\alpha}\}$ and $\{N_{\alpha\beta}^{\gamma}\}$~\cite{EntanglementBootstrap_long}.  Explicitly,
\begin{equation}
	d_n^2= \sum_{\alpha} N_{\mu(n)}^{\alpha} d_{\alpha},\label{eq:fusion_parton_general}
\end{equation}
where $d_{\alpha}$ is the quantum dimension of a point excitation on the domain wall ($\alpha \in \calC_O$), which is determined according to $d_{\alpha}d_{\beta}= \sum_{\gamma} N_{\alpha\beta}^{\gamma} d_{\gamma}$, where $\alpha,\beta,\gamma \in \calC_O$. Here $\mu$ is an embedding map from the set of $N$-type parton sectors to another set of superselection sectors.\footnote{For the purpose of our discussion, the nature of these superselection sectors is unimportant; what is important is that the fusion space dimensions are integers.}

For example, one of the gapped domain wall types between the non-Abelian $S_3$ quantum double~\cite{Beigi2010} (on the $P$ side) and the toric code (on the $Q$ side) has $\calC_N=\{1,n\}$ and $\calC_U=\{1\}$. Here, we have $d_n=\sqrt{2}$ for the nontrivial $N$-type parton sector $n$. The value $d_n=\sqrt{2}$ can be derived from Eq.~(\ref{eq:fusion_parton_general}) as follows. This model has 6 point excitation types on the domain wall, \emph{i.e.,} $\calC_O$ contains $6$ elements. The quantum dimension of these objects are $d_{\alpha}\in \{1,1,1,1,2,2\}$, determined by the fusion space dimensions $\{N_{\alpha\beta}^{\gamma}\}$.
For the nontrivial parton sector $n$, $N_{\mu(n)}^{\alpha}=1$ for two choices of $\alpha \in \calC_O$, both having $d_{\alpha}=1$, whereas $N_{\mu(n)}^{\alpha}=0$ for other choices of $\alpha$. Therefore, $d_n^2=2$ which leads to $d_n= \sqrt{2}$. 

In fact, because $N_{\mu(n)}^{1}=1$ and $d_{\alpha}\ge 1$ generally, $\sqrt{2}$ is the smallest possible choice of the quantum dimension other than 1. Therefore, the quantum dimension of parton sectors we define are ``quantized" in the same manner as the quantum dimensions of anyons.

To summarize, the definition proposed in Eq.~\eqref{eq:quantum_dimension_parton} obeys a set of constraints that are analogous to Eq.~\eqref{eq:fusion_anyon}. These constraints justify our definition.

\bibliography{ref}
\bibliographystyle{apsrev}

\end{document}